\documentclass[aps,prl,amssymb,superscriptaddress,footinbib]{revtex4}

\usepackage{amsthm,amsmath}
\usepackage[utf8]{inputenc} 

\usepackage{amsfonts}
\usepackage{graphicx}
\usepackage{epsfig}
\usepackage{subfigure}
\usepackage{psfrag}
\usepackage{amsbsy}
\newcommand{\ba}{\begin{align}}
\newcommand{\ea}{\end{align}}
\newcommand{\be}{\begin{equation}}
\newcommand{\ee}{\end{equation}}
\newcommand{\bea}{\begin{eqnarray}}
\newcommand{\eea}{\end{eqnarray}}

\usepackage{amsmath}
\usepackage{amssymb}
\usepackage{amsfonts}
\usepackage{color}
\usepackage{soul}
\usepackage{ulem} 
\usepackage{marginnote}
\usepackage{tikz}
\usepackage{dsfont}
\usepackage{hyperref}

\begin{document}

\title{Thermal noise in BEC-phononic gravitational wave detectors}

\author{Carlos Sab\'in}
\affiliation{Instituto de F\'isica Fundamental,
CSIC,
Serrano 113-bis,
28006 Madrid, Spain}

\author{Jan Kohlrus}
\affiliation{School of Mathematical Sciences,
University of Nottingham,
University Park,
Nottingham NG7 2RD,
United Kingdom}

\author{David Edward Bruschi \footnote{Current affiliation: York Centre for Quantum Technologies, Department of Physics, University of York, Heslington, YO10 5DD York, UK.}}
  \affiliation{Racah Institute of Physics and Quantum Information Science Centre, the Hebrew University of Jerusalem, 91904 Jerusalem, Israel}

\author{Ivette Fuentes}
  \affiliation{School of Mathematical Sciences,
University of Nottingham,
University Park,
Nottingham NG7 2RD,
United Kingdom}
\affiliation{University of Vienna, Faculty of Physics, Boltzmanngasse 5, 1090 Wien, Austria}

 \begin{abstract} Quasiparticles in a Bose-Einstein condensate are sensitive to space-time distortions. Gravitational waves can induce transformations on the state of phonons that can be observed through quantum state discrimination techniques.  We show that this method is highly robust to thermal noise and depletion. We derive a bound on the strain sensitivity that shows that the detection of waves in the kHz regime is not significantly affected by temperature in a wide range of parameters that are well within current experimental reach.\end{abstract}

\maketitle

\begin{figure}[t]
\includegraphics[width=0.5\textwidth]{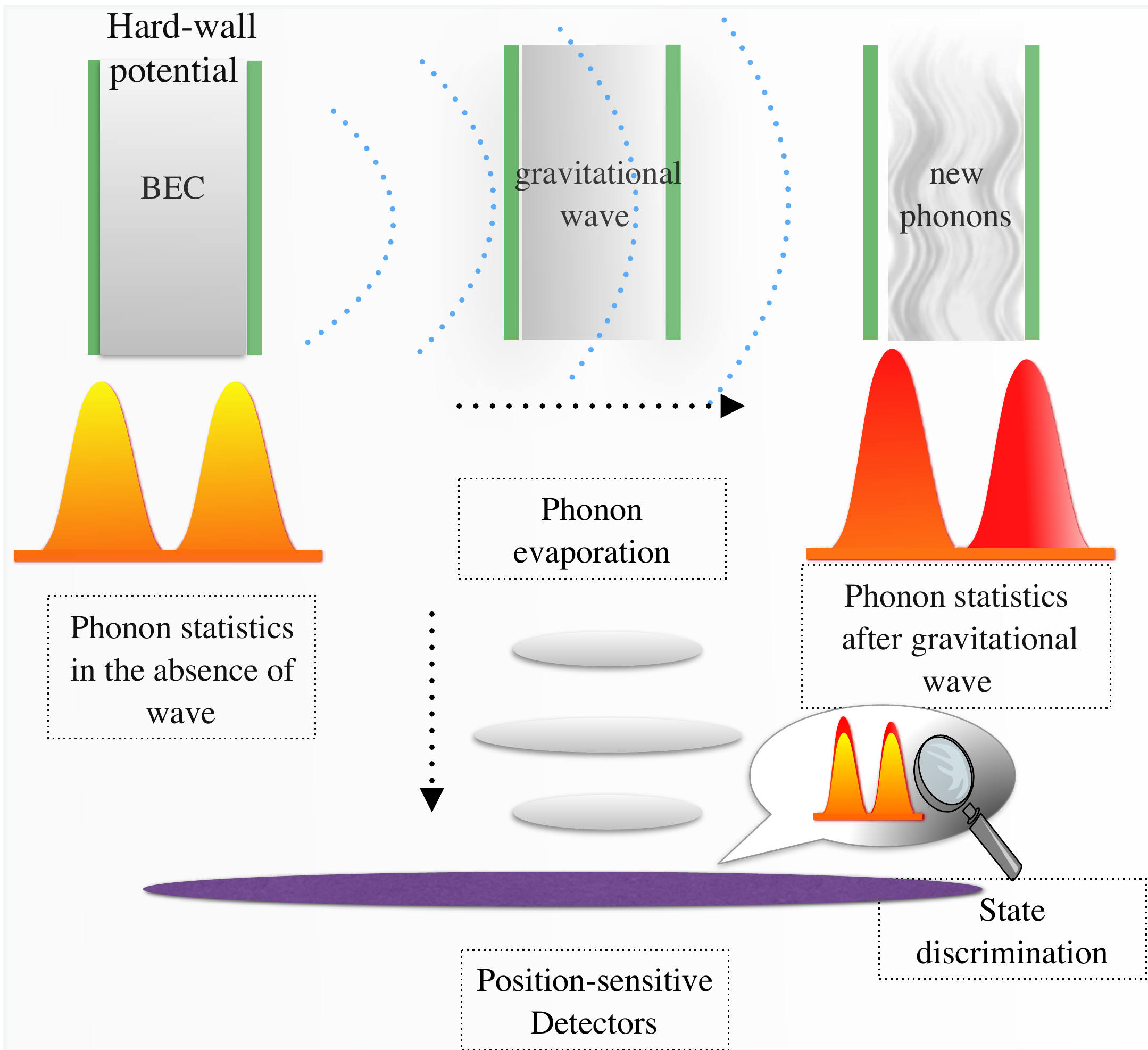}
\caption{Sketch of the setup: A BEC in a box-like potential that acts as a cavity for the phononic excitations. Two phononic modes $m$ and $n$ are initially prepared in a thermal two-mode squeezed state represented by the initial covariance matrix $\sigma_0$. A gravitational wave of amplitude $\epsilon$ transforms the state producing excitations. The new state $\sigma_{\epsilon}$ depends on $\epsilon$. Measurements on the modes can be used to estimate the amplitude of the spacetime distortion.}\label{fig:fig0}
\end{figure}

The detection of gravitational waves \cite{gravwavesdetectors} remains an open problem and represents one of the most ambitious enterprises of science in the 21st century. After years of active efforts \cite{gravwaveastronomy,quantumgravobs}, several large-scale experiments are still in operation around the globe, based both in laser interferometry -such as advanced LIGO, GEO 600 or VIRGO- and Weber bar detectors -such as AURIGA and Mario Schoenberg. However, no successful observation of a gravitational wave has been reported yet. Therefore, together with new upgrades of the existing setups, new major international projects are expected to start operations in the short and medium term, including space-based laser interferometers such as DECIGO and LISA. This enormous investment of resources is backed up by indirect evidences of the existence of gravitational waves as well as a number of experiments confirming the predictions of Einstein's General Relativity, a theory from which the existence of spacetime ripples is a natural consequence \cite{reviewgravwaves}. However, since the Earth is very far from typical sources of gravitational waves, the intensity of the latter is so tiny when it reaches our detectors that gravitational wave detection is always a daunting task. \footnote{After the submission of this work, a first event of gravitational wave detection was reported by the LIGO collaboration, which has been heralded as the opening of a new era in gravitational wave astronomy.}

Recently, a new way of detecting spacetime distortions was proposed using a different physical principle \cite{ourgwd}. The state of the quasiparticles of a Bose-Einstein condensate (BEC) is modified by the passing of the gravitational wave. If the frequency of the gravitational wave matches the sum of the frequencies of two BEC modes the transformation of the state is resonantly enhanced in a phenomenon resembling the Dynamical Casimir Effect \cite{moore,casimirwilson}, characterised by a linear growing in time of the transformed state. Indeed, the scheme is equivalent to an artificial modulation of the length of the BEC trap, which can be implemented by a modulation of the atomic interaction strength \cite{finke, simulationgw}.This gravitational quantum resonance is absent in laser interferometers since the frequency of a gravitational wave is very far from the optical regime and is also different of the vibrational resonances in Weber bars. In \cite{ourgwd}, we showed that the sensitivity of the setup is low enough to, in principle, enable the detection of gravitational waves in a certain experimental parameter regime. 

In this work, we show that our scheme (see Fig. (\ref{fig:fig0})) for gravitational wave detection is highly robust against the effects of thermal noise. In \cite{ourgwd} it was assumed that the quasiparticle state is prepared in a particular pure state before the passing of the gravitational wave. Now we include an initial temperature and show that the sensitivity is not significantly affected in a wide regime of temperatures well within experimental reach. Thus, we give a step further in the analysis of the experimental feasibility of our scheme for gravitational wave astronomy. Following the same spirit, we also show that the spacetime ripple does not generate additional thermal depletion on the atomic bulk of the condensate, but only induces a phase shift. Indeed, this phase shift is much less sensitive to the action of the gravitational wave. However, there have been proposals to detect gravitational waves with atom interferometer setups, both Earth and space-based \cite{stanford, bongs}. These experiments would aim at detecting low-frequency gravitational waves, from $10^{-5}$ to $10^{-2}\, \operatorname{Hz}$. Our  setup operates in a completely different frequency range, ranging from $\operatorname{Hz}$ to $\operatorname{KHz}$ -the same frequency range as LIGO. Moreover, the sensitivity improves with the frequency while the sensitivity of LIGO is optimal in the range of $100 \operatorname{Hz}$ and decreases at higher frequencies. In particular, this implies that models of spacetime waves generated by the merging of neutron star binary systems into massive neutron stars and black holes go beyond LIGO's capabilities while being in principle within the reach of our proposal \cite{kentaip}. Detecting gravitational waves in the $\operatorname{kHz}$ regime would allow us to deepen our understanding of neutron stars by gathering information about their mass and radius. These parameters are necessary to describe the Neutron star state equation\cite{Andersson:2013} and also allow cosmologists to compute distances that are key in the study of the cosmological constant and dark matter \cite{kentaprd}. Other recent proposals of high-frequency gravitational-wave detectors can be found in \cite{tobar, hfgw}.

Let us explain our model and results in more detail.  We describe the BEC on a general spacetime metric following references \cite{matt, liberati,salelites}.  In the quantum hydrodynamic regime, the BEC phase is described by a mean field classical background $\Psi$ plus quantum fluctuations $\hat\Pi$ \cite{pethicksmith}.
The dynamics of $\Psi$ is governed by the Gross-Pitaievskii equation. At thermal energies much lower than the chemical potential, we can neglect the thermal depletion and assume that all the atoms remain in the ground state. Under these conditions, the effect of a gravitational wave on $\Psi$ is just to induce a phase shift \cite{phaseshiftgw}. We will discuss this in more detail below. By now, let us focus on the  dynamics of the quantum fluctuations. For length scales larger than the so-called healing length, $\Pi$ behaves like a phononic quantum field on a curved metric.  Indeed, as long as the fluctuations in density along the condensate are so small that we can neglect the quantum pressure term, the field obeys a massless  Klein-Gordon equation $\Box\hat\Pi=0$ where the d'Alembertian operator $\Box=1/\sqrt{-\mathfrak{g}}\,\partial_{a}(\sqrt{-\mathfrak{g}}\mathfrak{g}^{ab}\partial_{b})$ depends on an effective spacetime metric $\mathfrak{g}_{ab}$ -with determinant $\mathfrak{g}$- given by  \cite{matt,liberati,salelites}
\begin{equation}
\mathfrak{g}_{ab}=\left(\frac{n^2_0\,c_s^{-1}}{\rho_0+p_0}\right)\left[g_{ab}+\left(1-\frac{c_s^2}{c^2}\right)V_aV_b\right].
\end{equation}
The effective metric is a function of the real spacetime metric $g_{ab}$ -that in general may be curved- and
background mean field properties of the BEC such as the number density $n_0$, the energy density $ \rho_0$, the pressure $p_0$ and the speed of sound $c_s:=c\sqrt{\partial p/\partial\rho}$.  Here $p$ is the total pressure, $\rho$ the total density and $V_a$ is the so-called 4-velocity flow on the BEC, given by the gradient of $\Psi$. The pressure $p$ and the density $\rho$ differ from the their bulk counterparts only in a small linear perturbation. This description stems from the theory of linearised perturbations of fluids in a general relativistic background \cite{matt}, and thus is valid as long as the BEC can be described
as a quantum fluid, that is as long as it remains within the quantum hydrodynamic regime
\cite{liberati}. In the absence of background flows and considering a single spatial dimension, we obtain $V_a=(c,0)$.  In this case the effective metric reduces to
\begin{equation}
\mathfrak{g}_{ab}= \left(\frac{n^2_0\,c_s^{-1}}{\rho_0+p_0}\right) \left[g_{ab}+ \begin{pmatrix}c^2-c_s^2&0\\0&0\end{pmatrix}\right].
\end{equation}
In the absence of a gravitational wave, the real spacetime metric is $g_{ab}=\eta_{ab}$, where $\eta_{ab}=(-c^{2},1)$ is the flat two-dimensional Minkowski metric and the effective metric $\mathfrak{g}_{ab}$ is then a Minkowski-like metric with the speed of light being replaced by the speed of sound $c_s$. The solutions of the Klein-Gordon equation are then massless excitations propagating with the speed of sound $c_s$. Therefore, the frequency of the mode $\omega_k$ is given by the dispersion relation $\omega_k=c_s\,|\bf{k}|$, where $k$ is the mode's momentum.  This linear dispersion is valid as long as $\hbar\,k<<m_{0}\,c_s$, where $m_{0}$ is the mass of the BEC's atoms.

We consider that the BEC is contained in a 1-dimensional cavity trap. Therefore, we choose to impose close to hard-wall boundary conditions \cite{condensatebox1, condensatebox2,condensatebox3} that give rise to the spectrum,
$\omega_n=\frac{n\,\pi\,c_s}{L},$ where $L$ is the cavity length and $n\in\{1,2...\}$.

The phononic field $\Pi(t,x)$ is then quantised by associating creation and annihilation operators $a^{\dagger}_k$  and $a_k$ to the mode solutions of the Klein-Gordon equation in the effective metric \cite{pethicksmith,ourgwd}, and can be expanded as $ \Pi(t,x)= \sum_{k}\,[\,\phi_{k}(t,x)\,a_{k}\,+\,\phi^{*}_{k}(t,x)\,a^{\dagger}_{k}]$. The bosonic operators $a_k$ and $a^{\dagger}_k$ obey the canonical commutation relations.

By restricting  our analysis to phononic Gaussian states, we are able to use the covariance matrix formalism to describe the dynamics of $\Pi$. Gaussian states of bosonic fields and their transformations take a very simple form in the covariance matrix formalism. This simplifies the application of quantum metrology techniques to relativistic quantum fields \cite{rqm,rqm2}. Considering a collection of $N$ bosonic modes, we define the quadrature operators $X_{2n-1}=\frac{1}{\sqrt{2}}(a_{n}+a^{\dag}_{n})$ and $X_{2n}=\frac{1}{\sqrt{2}\,i}(a_{n}-a^{\dag}_{n})$ where $n=1,\ldots,N$, which correspond to the generalised position and momentum operators of the field, respectively.  In the covariance matrix formalism Gaussian states are completely defined by the field's first and  second moments. Furthermore, quadratic linear unitary operators, such as Bogoliubov transformations, are represented by symplectic matrices $\boldsymbol{S}$ that satisfy $\boldsymbol{S}^T\,\boldsymbol{\Omega}\,\boldsymbol{S}= \boldsymbol{\Omega}$. Here, the matrix $\boldsymbol{\Omega}$ is the symplectic form defined by $\boldsymbol{\Omega}=\bigoplus_{k=1}^{n}\boldsymbol{\Omega}_k$, $\boldsymbol{\Omega}_k=-i\sigma_y$ and $\sigma_y$ is the corresponding Pauli matrix. The first moments of the state are $\langle X_{i}\rangle$ and the second moments are encoded in the covariance matrix $\boldsymbol{\sigma}$ defined by $\sigma_{ij}=\langle X_{i} X_{j}+X_{j}X_{i}\rangle-2\langle X_{i}\rangle\langle X_{j}\rangle$.
Without loss of generality, we restrict our analysis to initial Gaussian states with vanishing first moments, i.e. $\langle X_{i}\rangle=0$.

In this work we will consider the following initial state of the field $\boldsymbol{\sigma}_0$:
\begin{equation}
\boldsymbol{\sigma}_0=\boldsymbol{S}^{T}\,\boldsymbol{\nu}\,\boldsymbol{S}
\end{equation}
where $\boldsymbol{\nu}=\operatorname{diag}\{\nu_n,\nu_n,\nu_m,\nu_m\}$ is called the Williamson form and $\nu_k=\operatorname{coth}(\frac{\hbar\omega_k}{2 k_B\,T})$ are the symplectic eigenvalues of the initial state $\boldsymbol{\sigma}_0$ (i.e., the eigenvalues of the matrix $|i\boldsymbol{\Omega}\,\boldsymbol{\sigma}_0|$).
We will consider the quantum regime $\hbar\,\omega_k>>k_B\,T$ so we can expand $\nu_k$ as follows:
\begin{equation}\label{eq:initialstate}
\operatorname{coth}(\beta_k)=1+2\,e^{-2\beta_k}+\mathcal{O}(e^{-4\beta_k}).
\end{equation}
where $\beta_k:=\frac{\hbar\omega_k}{k_B\,T}\gg1$.

For relatively high frequencies, such as $\omega_1=2 \pi\times 5\cdot10^3\,\operatorname{Hz}$- which corresponds to $L=1\operatorname{\mu\,m}$ and $c_{s}=10\,\operatorname{mm/s}$- the above condition entails that we can consider temperatures up to $150\operatorname{nK}$ -which corresponds to $e^{2\,\beta}\simeq10$. This is a relatively high temperature for a BEC where $T$ can be even lower than $1\,\operatorname{nK}$ \cite{sciencewr, jeffblackhole}. Note however that we should be cautious in extending our analysis beyond a few $nK$, since we are neglecting thermal depletion in the condensate bulk, that is temperature is much smaller than the chemical potential $k_{B}\,T<<\mu$. For typical values of the chemical potential $\mu/k_B>100\,\operatorname{nK}$, which implies that it is reasonable to consider temperatures as large as $10\,\operatorname{nK}$. As the temperature grows, the effect of the thermal cloud on the dynamics of the quantum field modes might become relevant via the Beliaev damping mechanism \cite{jeffbeliaev}, which we are not considering here.

Note that if $m>n$ then,
$\beta_m=m\,\beta_n/n$. This means that
for $m=2$ and $n=1$, it is possible to neglect $e^{-\beta_m}$ when compared to $e^{-\beta_n}$. In the regime of temperatures considered here, we obtain $\boldsymbol{\nu}=\mathds{1}+\boldsymbol{\nu}^{(2)}$ and therefore, we can write the initial state as
\begin{equation}
\boldsymbol{\sigma}_0=\boldsymbol{S}^{T}\,(\mathds{1}+2\,\boldsymbol{\nu}^{(2)})\,\boldsymbol{S},
\end{equation}
where the small contribution $\boldsymbol{\nu}^{(2)}$ takes the form $\boldsymbol{\nu}^{(2)}=\text{diag}\{x_n ,x_n,x_m, x_m\}$ and $x_k=e^{-\beta_k}$. We will also consider that the transformation encoded in $\boldsymbol{S}$ is a two-mode squeezing transformation. This transformation can be implemented in the laboratory, for instance, by artificially modulating the length of the trap \cite{casimirwestbrook}, or equivalently by modulating the interaction strength \cite{finke, simulationgw}. However, the degree of squeezing achievable by this method would be limited mainly by depletion and losses, so perhaps alternate methods must be considered. Following \cite{ourgwd}, the frequencies of the squeezed modes resonate with the frequency of the gravitational wave, which in turn tunes our detector to a particular frequency bandwidth.

Under the action of the gravitational wave the real spacetime metric $g_{\mu \nu}$ is transformed from the Minkowski flat spacetime metric to a spacetime including a small perturbation $h_{\mu\nu}$, i.e. $g_{\mu\nu}=\eta_{\mu\nu}+h_{\mu\nu}$ (see \cite{reviewgravwaves}).
In one spatial dimension and the traceless-transverse (TT) gauge a perturbation corresponding to a gravitational wave moving in a transverse direction can be written as,
\begin{equation*}
h_{\mu\nu}= \begin{pmatrix}0&0\\0&h_{+}(t)\end{pmatrix},
\end{equation*}
where $h_{+}(t)$ is typically modelled as a sinusoidal function $h_{+}(t)=\epsilon\operatorname{sin}\Omega\,t$,
and $\epsilon$ and $\Omega$ are the amplitude and frequency of the spacetime ripple, respectively.
The change of the real spacetime metric $g_{\mu \nu}$ induces a change of the effective metric $\mathfrak{g}_{ab}$. This in turn generates a  Bogoliubov transformation $\boldsymbol{S}$ on the quantum field. The flat- spacetime field operators ${a_{k}}$  are transformed into,
$\hat{a}_{k}=\sum_{j} \bigl(\alpha^{*}_{kj}a_{j}+\beta^{*}_{kj}a^{\dag}_{j}\bigr)\,,$
where $\hat{a}^{\dagger}_k$ and $\hat{a}_k$ are creation and annihilation operators associated to the mode solutions in the perturbed spacetime and $\alpha_{kj}(h_+(t))$ and $\beta_{kj}(h_+(t))$ are Bogoliubov coefficients that depend on the wave's spacetime parameters. They were computed in \cite{ourgwd} in the case of a box-like BEC trap under the following assumptions.

We first consider that beam-pointing laser noise is negligible in the range of $\operatorname{kHz}$  frequencies \cite{savard,kwee,ohara} and thus the trap can be considered as rigid, and the intensity of the gravitational wave is so small that it remains rigid under its action. Second, the frequency of the gravitational wave matches the sum of two modes of interest $\Omega=\omega_m+\omega_n$. Third, the BEC-wave interaction time $t$ is long enough $\omega_1\,t>>1$. The latter two conditions hold for typical sources of gravitational waves. Under all the above conditions, the effect of the spacetime ripple is equivalent to a two-mode squeezing operation characterised by the coefficient $\beta_{mn}$, which grows linearly in time in a phenomenon resembling the Dynamical Casimir Effect.

The Bogoliubov transformation generated by the gravitational wave is encoded in the symplectic matrix $\boldsymbol{S}_\epsilon$ and the final state after the transformation is 
\begin{equation}\label{eq:state}
\boldsymbol{\sigma}_\epsilon=\boldsymbol{S}^{T}_\epsilon\,\boldsymbol{\sigma}_0\,\boldsymbol{S}_\epsilon,
\end{equation}
where
\begin{align*}
\boldsymbol{S}_\epsilon=\left(
  \begin{array}{cccc}
    \boldsymbol{\mathcal{M}}_{11} &  \boldsymbol{\mathcal{M}}_{12} &  \boldsymbol{\mathcal{M}}_{13} & \cdots \\
     \boldsymbol{\mathcal{M}}_{21} &  \boldsymbol{\mathcal{M}}_{22} &  \boldsymbol{\mathcal{M}}_{23} & \cdots \\
     \boldsymbol{\mathcal{M}}_{31} &  \boldsymbol{\mathcal{M}}_{32} &  \boldsymbol{\mathcal{M}}_{33} & \cdots \\
    \vdots & \vdots & \vdots & \ddots
  \end{array}
\right)
\end{align*}
and the $2\times2$ matrices $ \boldsymbol{\mathcal{M}}_{mn}$ are given by
\begin{align*}
\boldsymbol{\mathcal{M}}_{mn}=\left(
                   \begin{array}{cc}
                     \Re(\alpha_{mn}-\beta_{mn}) & \Im(\alpha_{mn}+\beta_{mn}) \\
                     -\Im(\alpha_{mn}-\beta_{mn}) & \Re(\alpha_{mn}+\beta_{mn})
                   \end{array}
                 \right)\,.
\end{align*}
Here $\Re$ and $\Im$ denote the real and imaginary parts, respectively. The Bogoliubov coefficients and thus, the final state of the field, depend on the amplitude of the gravitational wave $\epsilon$.

The main goal of this work is to analyse the impact of the temperature $T$ in the bound on the optimal precision that can be achieved when estimating the wave amplitude $\epsilon$ through measurements on the state $\boldsymbol{\sigma}_\epsilon$. The quantum Cramer-Rao theorem states that the error in the estimation of the parameter $\epsilon$ is bounded by $\langle (\Delta \hat{\epsilon})^{2}\rangle\geq \frac{1}{MH_\epsilon}$, where $H_{\epsilon}$ is the Quantum Fisher Information (QFI) and $M$ the number of probes. The QFI can be computed using the Uhlmann fidelity~$\mathcal{F}$ between the state $\boldsymbol{\sigma}_{\epsilon}$ and a state $\boldsymbol{\sigma}_{\epsilon+d\epsilon}$ with an infinitesimal increment in the parameter. In particular one has $H_\epsilon=\frac{8\big(1-\sqrt{\mathcal{F}(\boldsymbol{\sigma}_{\epsilon},\boldsymbol{\sigma}_{\epsilon+d\epsilon})}\big)}{d\epsilon^{2}}$.

Now let $\boldsymbol{\sigma}_{\epsilon}$ be a two-mode Gaussian state with zero initial first moments. The fidelity is then given by
\cite{MarianMarian}
\begin{align}\label{uhlmann:fidelity}
    \mathcal{F}(\boldsymbol{\sigma}_{\epsilon},\boldsymbol{\sigma}_{\epsilon+d\epsilon})=\frac{1}{\sqrt{\Lambda}+\sqrt{\Gamma}-\sqrt{(\sqrt{\Lambda}+\sqrt{\Gamma})^2-\Delta}},
\end{align}
where we have introduced
$\Gamma =\,\frac{1}{16}\text{det}(i\mathbf{\Omega}\,\boldsymbol{\sigma}_{\epsilon}\,i\mathbf{\Omega}\,\boldsymbol{\sigma}_{\epsilon+d\epsilon}+1_{2\times2})$,
$\Lambda =\,\frac{1}{16}\text{det}(i\mathbf{\Omega}\,\boldsymbol{\sigma}_{\epsilon}+1_{2\times2})\,\text{det}(i\mathbf{\Omega}\,\boldsymbol{\sigma}_{\epsilon+d\epsilon}+1_{2\times2})$ and $\Delta =\,\frac{1}{16}\text{det}(\boldsymbol{\sigma}_{\epsilon}+\boldsymbol{\sigma}_{\epsilon+d\epsilon})$.
In \cite{rqm2}, it is shown that in the case in which the initial state $\boldsymbol{\sigma}_0$ is pure, $\Gamma=\Delta+\mathcal{O}(\epsilon^4)$ and $\Lambda=\mathcal{O}(\epsilon^2(\epsilon+d\epsilon)^2)$ and so the only relevant contribution to the QFI comes from the computation of $1/\sqrt{\Gamma}$. In the case that we are analysing here, namely a quasi-pure initial state where temperature is only a small perturbation, we find that this conclusion is still true. To lowest order, it is possible to show that $\Gamma=\Delta+\mathcal{O}((x_n+x_m)(\epsilon+d\epsilon)^2)$ and $\Lambda=\mathcal{O}((x_n+x_m)^2(\epsilon+d\epsilon)^4)$. This implies that the total contribution of the second square root of \eqref{uhlmann:fidelity} cannot provide the terms proportional to $dh^2$ which are the terms that contribute to the final QFI. Furthermore, since $\Lambda=\mathcal{O}((x_n+x_m)^2(\epsilon+d\epsilon)^4)$, one concludes that $\sqrt{\Lambda}$ does not contribute to the final $dh^2$ term either. Therefore, the only contribution to the QFI is given by $\Gamma$, as argued before. If one wishes to find at which order the temperature will contribute, one needs to compute higher order corrections to all terms in the fidelity. It is possible to show, however, that the temperature will contribute \textit{linearly} to the zero order QFI with a term proportional to $x_n,x_m$.
\begin{figure}[h!]
\includegraphics[width=0.5\textwidth]{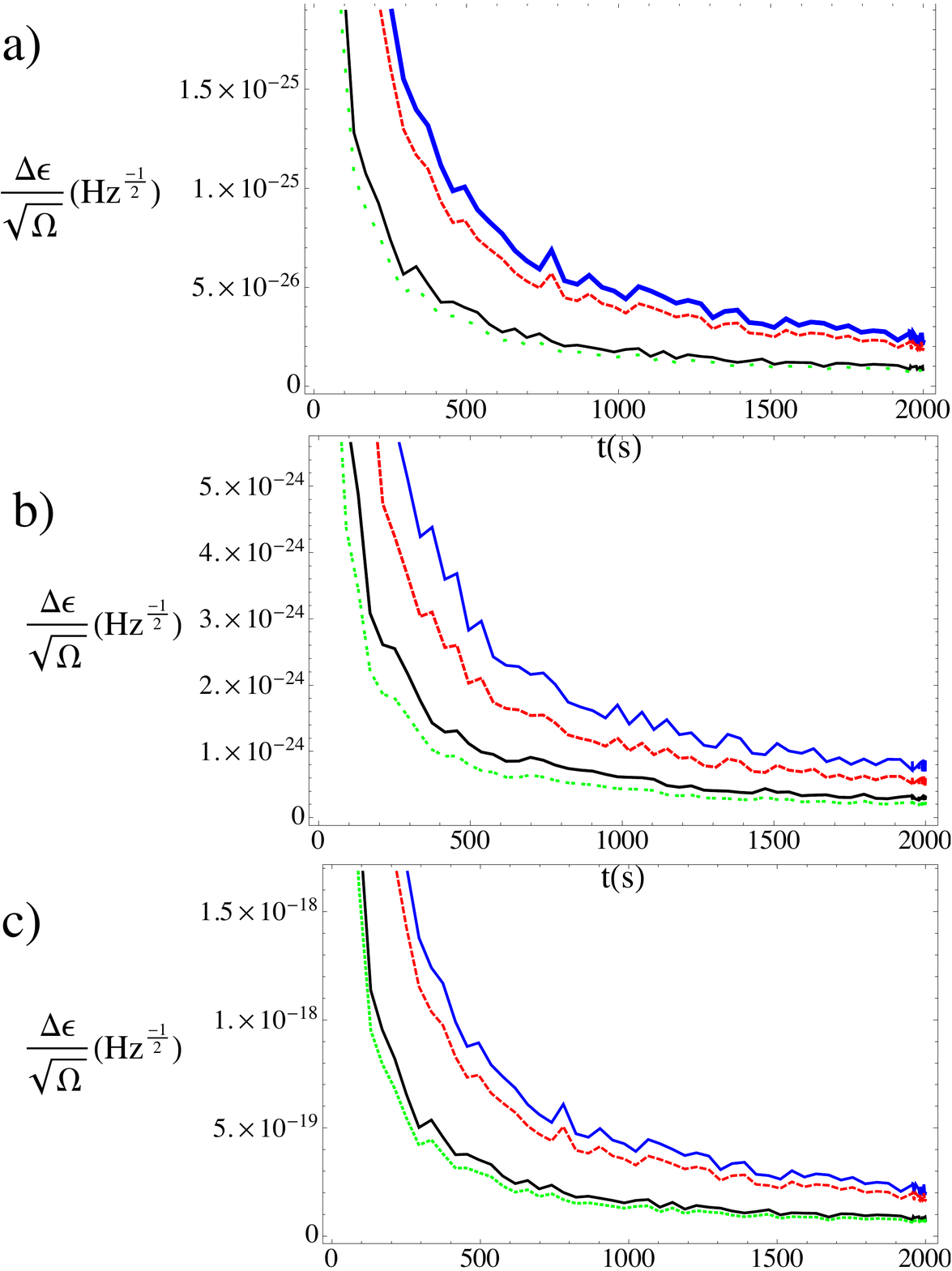}
\caption{Time evolution of the optimal bound for the strain sensitivity as provided by the QFI for  $n=1$, $m=2$ ,  $T=0$ (blue, solid), $T=150\, \operatorname{nK}$ (red,dashed),  $m=6$ ,  $T=0$ (black, solid), $T=150  \operatorname{nK}$ (green,dotted) and a) $r=10$, $\omega_1=5\,10^3\operatorname{Hz}$, b) $r=10$, $\omega_1=5\,10^2\operatorname{Hz}$, c) $r=2$, $\omega_1=5\,10^3\operatorname{Hz}$. \label{fig:results}}
\end{figure}

\begin{figure}[h!]
\includegraphics[width=0.5\textwidth]{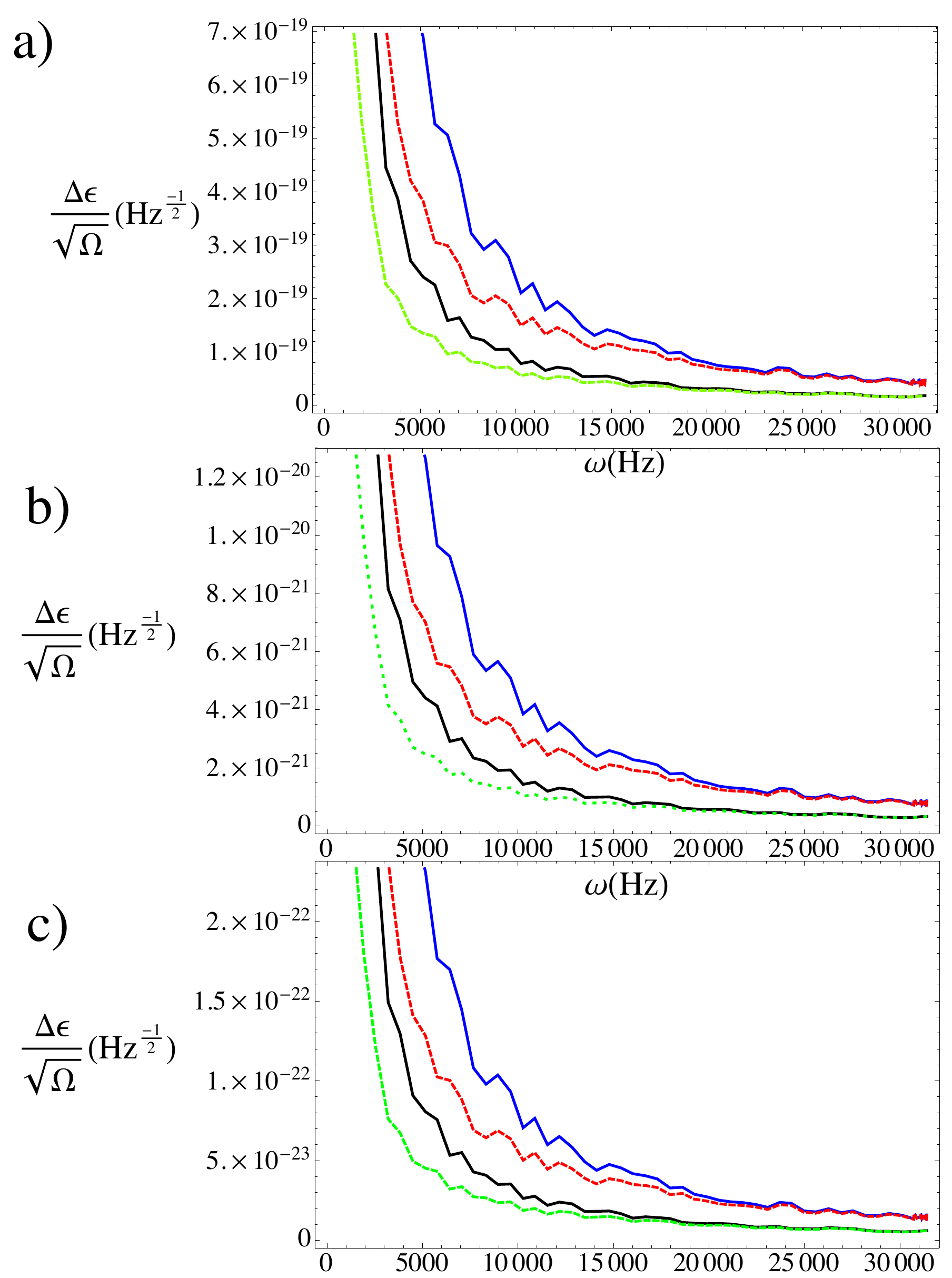}
\caption{Optimal bound for the strain sensitivity as provided by the QFI vs frequency for  $t=10$,$n=1$, $m=2$ ,  $T=0$ (blue, solid), $T=150\, \operatorname{nK}$ (red,dashed),  $m=6$ ,  $T=0$ (black, solid), $T=150  \operatorname{nK}$ (green,dotted) and a) $r=5$,  b) $r=7$,  c) $r=9$. \label{fig:results2}}
\end{figure}

We proceed to illustrate numerically what we described above. By using Eqs. (\ref{eq:initialstate}), (\ref{eq:state}) and (\ref{uhlmann:fidelity}) and taking into account the discussion below the latter, we compute the QFI $H_{\epsilon}$. Since the analytical computations are cumbersome we show here plots obtained by numerical means. In Fig.\ref{fig:results} we show that the the optimal bound for the strain sensitivity provided by the QFI is very robust to initial temperatures. Actually, it is even improved at relatively high temperatures -a fact that is related with the already well-known increase of quantum correlations with temperature \cite{gaussip}. However, note that strictly speaking at those high temperatures the dynamics of the condensate would be different and should include damping mechanisms that we are not considering here \cite{richard}. The only reason to consider temperatures as high as $150\,\operatorname{nK}$ is to illustrate the fact that at lower temperatures the plots would be indistinguishable from the ones at $T=0$. Thus, restricting ourselves to the region $k_B\,T<<\mu$, where we can neglect thermal depletion in the atomic cloud, we conclude that thermal noise in the phononic state does not affect significantly the performance of a BEC-phononic gravitational wave detector. A remarkable feature of our mechanism is that the sensitivity improves in the $\operatorname{kHz}$ regime, as can be seen in Fig.\ref{fig:results2}. As discussed in the introduction, this would enable the test of gravitational waves coming from binary mergers, which are now beyond the reach of current detectors. Finally,  Fig. \ref{fig:results}  shows that the correction to the vacuum results is, to very good approximation, linear (corresponding to a translation of the plot). This corroborates the theoretical analysis which finds that the corrections to the QFI for small temperatures corresponds to a small correction to the vacuum QFI.

Let us now review the validity of some underlying approximations. We have considered that thermal depletion is negligible in the atomic bulk and is not significantly modified by the action of the gravitational wave. Indeed, in the case of negligible thermal depletion, the wave function of the condensate would acquire a phase $\Psi(t)=-\hbar\,k^2/(2\,m_{0})\,(t-\epsilon\cos{\Omega\,t}/\Omega)$, where in a 1D box-like potential $k=\pi\,c/L$ and $m_{0}$ is the mass of the BEC's atoms \cite{phaseshiftgw}. Since the velocity flows are defined by $V^{\mu}=c\,u^\mu/||u||$, where $u^\mu=\hbar/m_{0}\,\partial^\mu\Psi$ \cite{liberati}, this means that $V=(c,0)$ both in the absence of the gravitational wave and under its action -in agreement with our assumptions. Moreover, since the spacetime ripple only generates a phase shift, it does not change the number of condensed and depleted atoms. So if the condensate is initially prepared in a state where thermal depletion is negligible, the thermal depletion would remain negligible when the condensate undergoes interaction with the wave. Finally, it is interesting to analyse whether the phase shift of the condensate bulk could be used in order to detect the spacetime ripple or not. Indeed the QFI $H_{\epsilon}$ associated to a quantum state $\phi (t)=\phi_0\,e^{i\Psi(t)}$ is given by $H_{\epsilon}=|\partial_\epsilon \Psi(t)|^2$. Therefore, in this case $H_\epsilon=(\hbar\,k^2/(2\,m_{0}\Omega)\cos{\Omega\,t})^2$. Thus, we first note that, unlike the quadratic growth in time of the QFI in our scheme, the QFI of the bulk would only oscillate in time. Furthermore, considering the mass of ${}^{87}Rb$ and the same values of $L$ and $\Omega$ that we are considering in the rest of the work, the maximum value of this atomic QFI is $H_{\epsilon}\simeq10^{-2}$, which would provide a bound for the strain sensitive of approximately $10^{-8}\,\operatorname{Hz^{-1/2}}$ with the same parameters of Fig. (\ref{fig:results}a). Therefore, the atomic phase shift is extremely less sensitive than our mechanism and cannot be used for gravitational wave detection.

\begin{figure}[h!]
\includegraphics[width=0.5\textwidth]{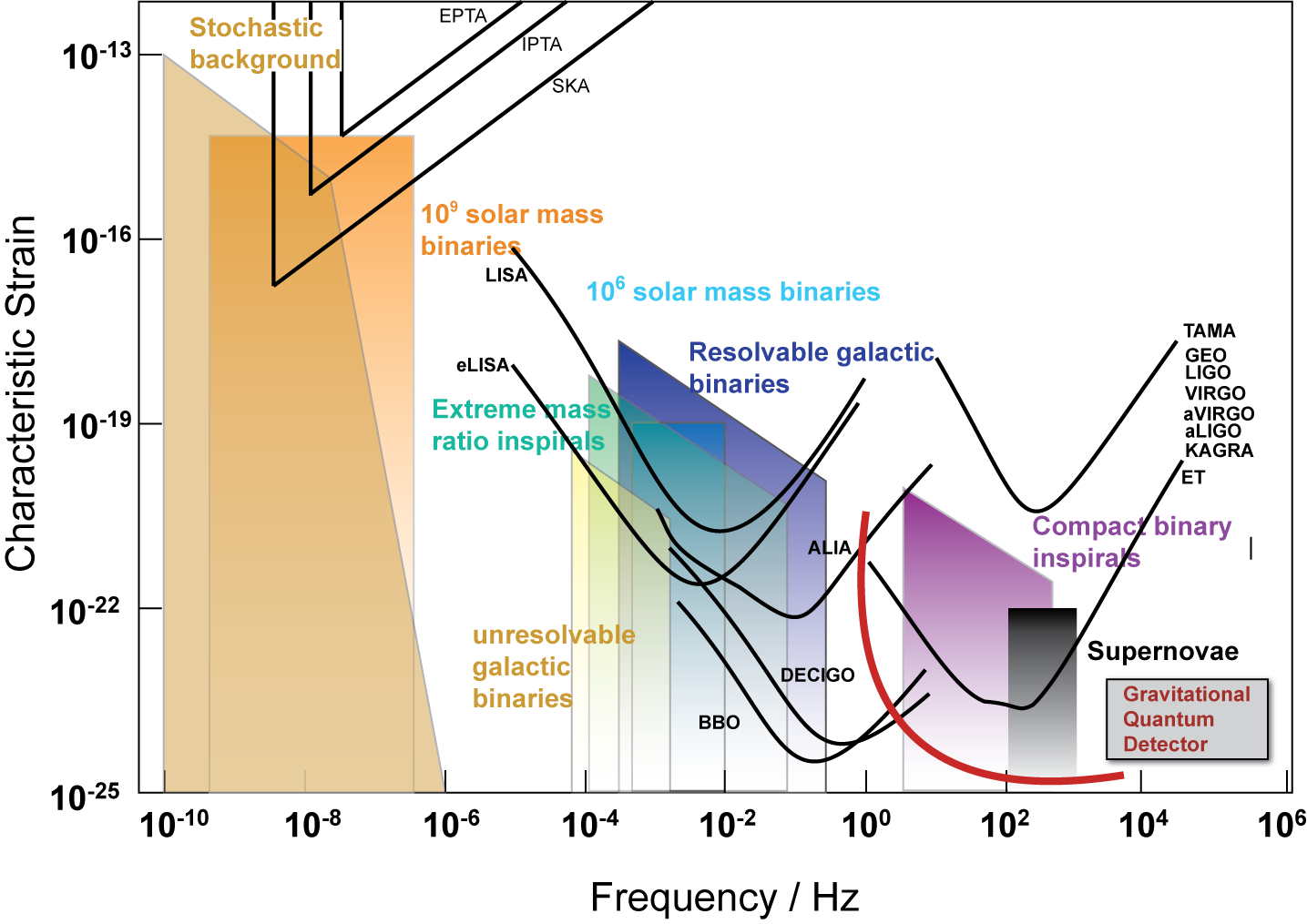}
\caption{Scaling of the strain sensitivity of our system (solid red line) as a function of frequency in comparison with other gravitational wave detectors (solid black lines). \label{Fig2_frequencies_v2}}
\end{figure}

Finally, in order to compare with the sensitivity of state of the art technology (i.e., LIGO, advanced LIGO, VIRGO and others), we plot the characteristic strain against the frequency together with that of other detectors. The results can be seen in Fig. (\ref{Fig2_frequencies_v2}).

Summarising, we have shown that the performance of our scheme for BEC-phononic gravitational wave detection is not significantly affected by the presence of initial thermal noise in the state of the phonons. The gravitational quantum resonance between the spacetime ripple and the quasiparticle modes in the BEC trap, is still present in a wide regime of temperatures, well within reach of cutting-edge cold-atoms technology. This represents a step further in the feasibility analysis of this novel scheme of gravitational wave astronomy, aiming to complement the open ambitious quest for gravitational waves.

\section*{Appendix}

Throughout this work, we have used the QFI as the figure of merit for sensitivity. The QFI provides an ideal bound, since it is obtained from an optimization of the  classical Fisher information over the set of all the possible measurements. However, the optimal measurement might not be easily implementable in the lab and the classical Fisher information of accessible measurements could be far from the ideal bound. This could question our choice of the QFI as a suitable figure of merit. However, in this Appendix we compute the classical Fisher information of a particular realistic measurement and show that it is close to the QFI in a relevant experimental parameter regime.

We consider that the estimation of the amplitude of the wave is achieved through the  measurement of the number of phonons in each mode. In \cite{casimirwestbrook}, this is achieved for modes carrying different momenta. This scheme relies on two key elements: first, a phonon evaporation process, according to which a phonon evolves to an atom of the same momentum when the BEC trap is released under adiabatic conditions. Second, after some free-fall time the atoms are collected on an array of position-sensitive detectors, from which is possible to infer the arrival time and initial velocity \cite{sciencewestbrook} -and thus the momenta of the phonons. The frequency range in \cite{casimirwestbrook} is the same that we are considering in this work -namely, $\operatorname{KHz}$. 

\begin{figure}[h!]
\includegraphics[width=0.5\textwidth]{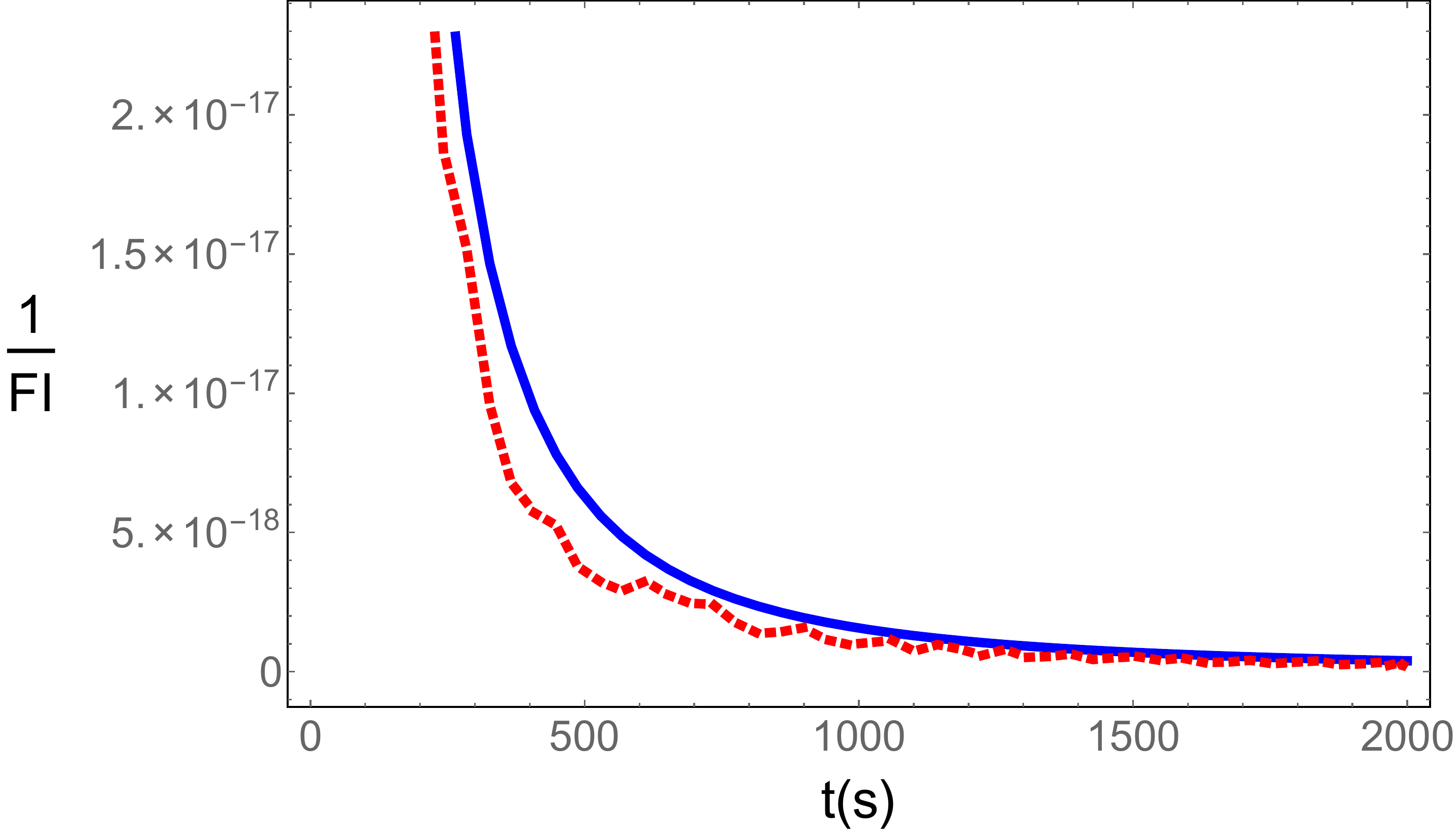}
\caption{Inverse of the classical (blue, solid) and quantum (red, dashed) Fisher information vs time for  $n=1$, $m=2$, $\omega_1=5 10^3 \operatorname{Hz}$,  $T=50\, \operatorname{nK}$ and $r=2$. \label{fig:fig4}}
\end{figure}

Then, the classical Fisher information is given by:
\begin{equation}
F_{\epsilon}=\sum_j\frac{1}{P(\epsilon|j)}\large|\frac{\partial P(\epsilon|j)}{\partial\epsilon}\large|^2,\label{eq:classfish}
\end{equation} 
where $P(\epsilon|j)$ is the probability distribution of the state, conditioned to the  result $j$ of the measurement- which in this case means the detection of $j$ phonons in each mode. In the case of the thermal two-mode squeezed state considered in this work, we have:
\begin{equation}
P(\epsilon|j)=\frac{1-e^{-\beta_n}-e^{-\beta_m}}{\cosh{\large(r+\beta_{nm}(\epsilon)\large)}}\tanh^j{\large(r+\beta_{nm}(\epsilon)\large)},\label{eq:probability}
\end{equation}
where all the magnitudes have been defined in the main text. Using Eqs. (\ref{eq:classfish}) and (\ref{eq:probability}) we can compute the classical Fisher information and compare it to the QFI. In Fig. (\ref{fig:fig4}) we see that the classical Fisher information is actually very close to the QFI for relevant experimental parameters. We have observed a similar behavior in all the parameter range explored in this work. Therefore, the use of the QFI as a figure of merit is well justified. 

\section*{Acknowledgments}
We thank Kenta Hotokezaka and Lucia Hackermuller for useful comments and discussions. D. E. Bruschi was supported by the I-CORE Program of the Planning and Budgeting Committee and the Israel Science Foundation (grant No. 1937/12), as well as by the Israel Science Foundation personal grant No. 24/12. Financial support by Fundaci{\'o}n General CSIC (Programa ComFuturo) is acknowledged by C. S.

 \end{document}